# MODELING OF THE OPTICAL PROPERTIES OF SILVER WITH USE OF SIX FITTING PARAMETERS

**A.O.Melikyan**

*Institute of Mathematics and High Technologies, Russian-Armenian (Slavonic) State University, Yerevan*
armen_melikyan@hotmail.com

and

**B.V.Kryzhanovsky**

*Departament of Nanotechnologies, Center of Optical Neural Technologies SRISA RAS, Moscow*
kryzhanov@gmail.com

**Abstract**  We propose a realistic model of the optical properties of of silver, in which inter-band transition with a threshold energy of ~ 4 eV is described phenomenologically by an ensemble of oscillators with same damping constant and a certain distribution of resonant frequencies in the interband transition threshold to infinity. The contribution of the conduction electrons in the dielectric function is determined by the Drude formula. The proposed model actually contains the features of both the Drude-Lorentz model (Rakić et al. 1998) and Tauc-Lorentz model (Jian-Hong Qiu et al. 2005). However, unlike these works proposed model contains only six fitting parameters, with the square root of the mean square deviation of the absorption coefficient and refractive index of silver from the experimental values in the range of 0.6 nm - 6.0 nm being of the order of 0.05.

## 1. INTRODUCTION

Modeling of the optical properties of various materials and parallel comparison of results with experimental data allows to identify the most important physical processes that determine the dependence of the dielectric function (DF) on frequency. Some progress in this area has been made for the noble metals, which are intensively investigated in connection with applications in nanooptics and photonics [1-5]. In the low-frequency region of the spectrum where the photon energy lies below the interband transitions threshold the behavior of the DF of silver is sufficiently well described by the Drude-like model in the range 1.8 ÷2.8 eV [4,5]. At the same time, at energies above the threshold of interband transitions good agreement is achieved for gold with the use of two or four damped oscillator model in addition to the free electron contribution. Another approach is based on the so called critical point model [1-3]. A universal approach using 17 parameters for each metal was developed in [6].

It is clear that the superposition of a sufficiently large number of Lorentz oscillators can describe properly the frequency dependence of the DF. However, the physical nature of the electron dispersion in the conduction band and the d-band sufficiently differs from the Lorentz oscillator physics. In this connection it should be noted that in [7, 8] a better agreement is achieved with use of the Tauc-Lorentz model, which is a finite ensemble of Lorentz oscillators with eigenfrequencies distributed continuously in the range from the threshold frequency to infinity. However in this model there is no Drude term describing the contribution of the conduction electrons, and an additional fitting parameter corresponding to the DF at asymptotically high photon energies is included.

In this paper we introduce a hybrid model, containing only six parameters in which the contribution of free electrons is described by the Drude formula with two adjustable parameters, and the contribution of the interband transitions by an ensemble of Lorentz oscillators with a continuous distribution of the eigenfrequencies and with the same damping constants. The density of states is chosen in such way that at electron energies close to the threshold of interband absorption the dispersion law is quadratic, and at high energies nonparabolicity of the conduction band is taken into account.

## 2. DESCRIPTION OF THE MODEL

Let $D(\omega_0)$ be the distribution function of the eigenfrequencies of the oscillator, then DF of silver in the proposed model can be represented as

$$\varepsilon(\omega) = 1 - \frac{\omega_p^2}{\omega(\omega + i\gamma_D)} + \int_0^\infty \frac{D(\omega_0)d\omega_0}{\omega_0^2 - \omega^2 - i\gamma\omega} \qquad (1)$$

where $\omega_p$ is the plasma frequency and $\gamma_D$ Drude relaxation constant.

We consider the contributions to the dielectric constant of silver $\varepsilon(\omega)$ from the plasma oscillations of s-electrons and the interband transition with the threshold energy of ~ 4 eV of d-electrons. For the former we use the simple Drude expression and the latter is the contribution of the interband transitions.

We choose $D(\omega_0)$ in the following form

$$D(\omega_0) = \begin{cases} f\sqrt{\omega_p\omega_0}\ e^{-\omega_0/\sigma}, & \omega_0 \geq \Delta \\ 0, & \omega_0 < \Delta \end{cases} \qquad (2)$$

where $f$, $\gamma$, $\Delta$ and $\sigma$ are fitting parameters of the interband transition: $\Delta$ is the interband transition threshold, $f$ is oscillator strength, and $\gamma$ is the damping constant. Choice of the partition function in the form as in expression (2) can be easily interpreted: if the partition function for the electron momentum is written as $W(p)dp \sim \exp[-p^2/p_0^2]p^2 dp$, then taking into account the dependence of electron energy on momentum and substituting $p^2 \to \omega_0$, $dp \to d\omega_0/\sqrt{\omega_0}$, $p_0^2 \to \sigma$ one can obtain $W(p)dp \to D(\omega_0)d\omega_0$. Finally we write down the expression for the DF with six fitting parameters as follows

$$\varepsilon(\omega) = 1 - \frac{\omega_p^2}{\omega(\omega + i\gamma_D)} + f\int_\Delta^\infty \frac{\sqrt{\omega_p \omega_0}\, e^{-\omega_0/\sigma} d\omega_0}{\omega_0^2 - \omega^2 - i\gamma\omega} \quad (3)$$

Obviously, the complex dielectric constant, defined by the expression (3) satisfies Kramers-Kronig dispersion relations.

To determine the constants in the Eq. (3) one should compare the results of simulation for the refractive index $n = n(\omega)$ or extinction coefficient $k = k(\omega)$

$$k(\omega) = \tfrac{1}{\sqrt{2}}\left[\sqrt{\varepsilon_1^2 + \varepsilon_2^2} - \varepsilon_1\right]^{1/2}$$
$$n(\omega) = \tfrac{1}{\sqrt{2}}\left[\sqrt{\varepsilon_1^2 + \varepsilon_2^2} + \varepsilon_1\right]^{1/2} \quad (4)$$

where $\varepsilon_1 = \mathrm{Re}\,\varepsilon(\omega)$ and $\varepsilon_2 = \mathrm{Im}\,\varepsilon(\omega)$, with the experimental data. It is known (see e.g. [9]) that the results of measuring of the absorption coefficient are more reliable as compared with the data on the refractive index. The data from the different sources such as [9-11] differ even in the long wavelength region, which is well described by the Drude term. At the same time, it turns out that the extinction coefficient is more sensitive to values of the parameters than the refractive index. For these reasons we carry out the fitting procedure using the data for the extinction coefficient $k = k(\omega)$. The values obtained by the least-squares method are given below:

$$\gamma_D = 0.022\ \mathrm{eV},\ \omega_p = 9.042\ \mathrm{eV},\ \Delta = 4.050\ \mathrm{eV}$$
$$\gamma = 0.260\ \mathrm{eV},\ \sigma = 9.935\ \mathrm{eV},\ f = 2.994 \quad (5)$$

It is important to note that the values of Drude parameters $\gamma_D$ and $\omega_p$ are determined from the best fit with the experimental data in the range from 0.64 eV to 1 eV, where the contribution of the interband transition can be neglected.
It is found that the root-mean-square deviation of $\delta k(\omega)$ value for the frequency range $0.64 \div 6.22\,\mathrm{eV}$ is equal to 0.067. Calculations for the refractive index give close values.

Figure 1 show respectively the results for real and imaginary parts of dielectric function, the refractive index and extinction coefficient according to the expressions (3)-(4) (solid lines) with use of the values of (5). The interpolation curves for the same quantities according to the measured data are shown by dotted lines.

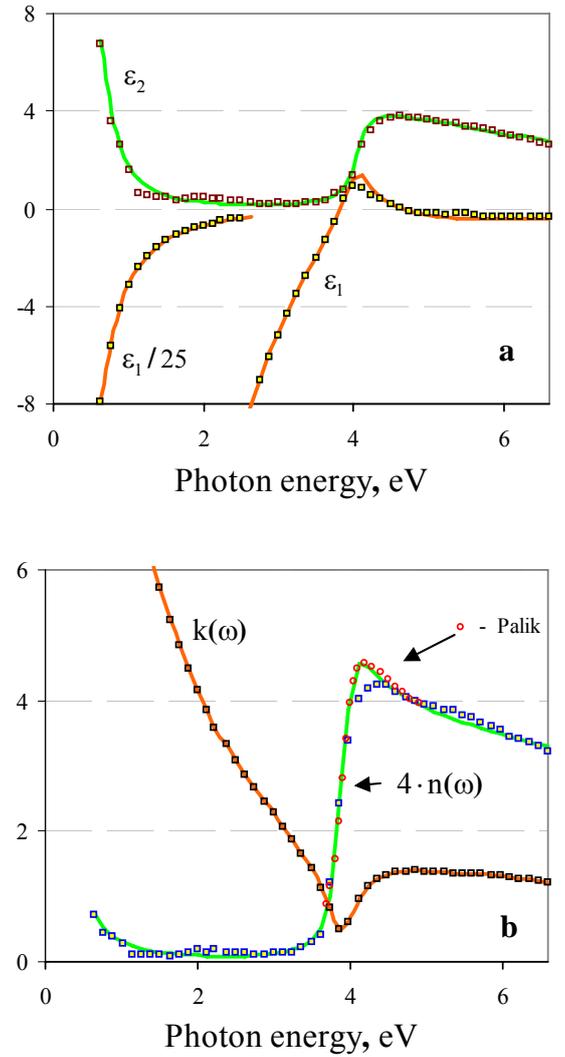

Fig.1. Real and imaginary parts of the optical dielectric function of Ag (a) and the refractive index $n = n(\omega)$ and extinction coefficient $k = k(\omega)$ of Ag (b). Solid lines correspond to the expressions (3) and (4); dotted lines are the experimental data points from Jonson and Christy [9] and for $\hbar\omega \in [3.5\,\mathrm{eV}, 4.5\,\mathrm{eV}]$ from Palik [10]. The values of the fitting parameters are given in (5).

### 3. RESULTS AND DISCUSSION
First, it can be seen that the results of calculations using the expressions (3) and (4) manifest excellent agreement with the experimental data. In addition, the values of the Drude parameters $\omega_p$ and $\gamma_D$ estimated by the least square method are very close to those extracted from the experimental data (see [9]). It is interesting to note that the value of $\gamma_D$ for Ag determined in [9] from the optical measurements agrees with that extracted from the measurement of conductivity [12].

Note that (3) describes adequately the optical properties of silver only at frequencies below 6 eV. At high frequencies, the new terms should be added similar to the integral in (3) describing the contribution of interband transitions at higher frequencies. The analysis shows that the contribution of the next interband transition (with a peak at ~ 14 eV [13]) becomes significant (~ 10%) even at frequencies close to 6 eV.

In Fig. 2 we present the results of calculations based on (3)-(5) for the transmission T and reflection R coefficients of silver films of thickness 20 and 50 nm on a quartz substrate. Solid lines correspond to the transmittance and reflectance of films with thickness of 20 and 50 nm, respectively, calculated with use of (3) – (5), while the squares and diamonds correspond to the data of [9].

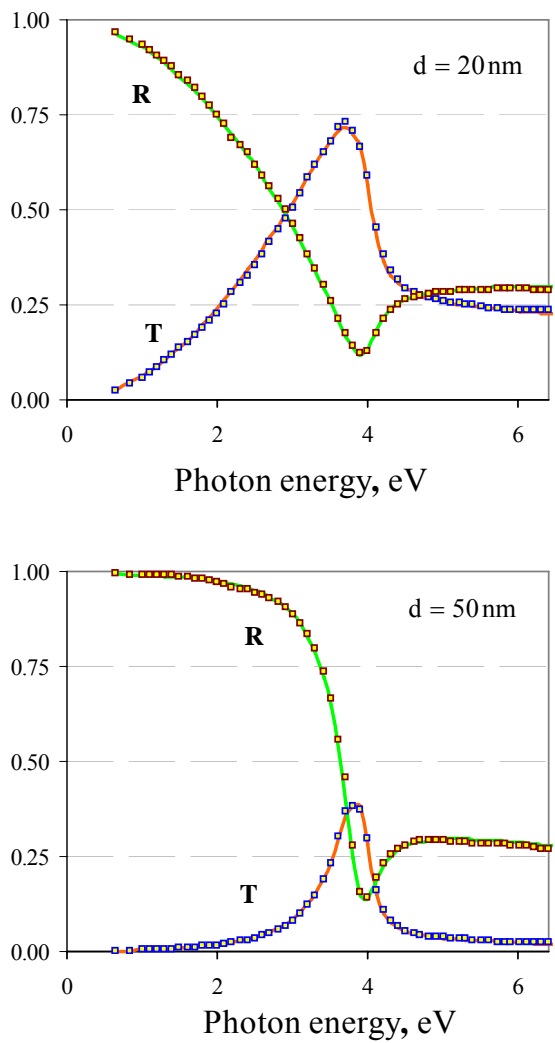

Fig 2. Transmission $T$ and refraction $R$ coefficients of silver films of thickness 20 nm (a) and 50 nm (b) on quartz substrate. Solid lines correspond to the transmittance and reflectance calculated with use of (3) –(5), squares and diamonds correspond to the data points of [9].

It should be noted that the fitting procedure based on the expression (3) with use of the SOPRA data [11] leads to the worst match even at low frequencies. In particular it gives for the relaxation constant $\gamma_D$ the value of 0.085 which is four times more than the value of [9]. This discrepancy in the data is easily explained. The matter is that the measurement of the absorption coefficient is carried out indirectly - on the transmission of a thin film. A transmission (extinction) is determined by the absorption in the film and the scattering by inhomogeneities, which is difficult to measure and, moreover, to account for. For this reason deposition of thin silver film is recommended to perform at low (less than 150°C) temperature of the substrate in order to minimize a film crystallite size and resulting dispersion. (for Au see [9], for Ag - [14]). The presence of impurities should be accounted for as well. It is especially difficult to get rid of the impurities of copper, which, even in the case of making a film of high purity silver (~ 99.99) can make a significant contribution to the measured value of the imaginary component of the dielectric constant. For example on the dispersion curve $\varepsilon_2 = \mathbf{Im}\, \varepsilon(\omega)$ shown in [9], a flat maximum in the frequency range $\omega \sim 2\,\text{eV}$ is clearly visible which corresponds to the maximum (with the value of ~ 6) of the imaginary part of the dielectric function of copper.

## 4. CONCLUSION

Realistic phenomenological model of the permittivity of silver as a function of frequency is proposed that takes into account qualitatively the band structure. The values of six fitting parameters are determined using the least-square method and the experimental data of [9]. The model operate with only six parameters for Drude term and one interband transition with threshold energy of about 4 eV, and provides excellent agreement with the data of [9] in the range of photon energy from 0.62 eV to 6.22 eV. At the same time more complicated model [6] in which 21 fitting parameters are used gives very large deviations from the experimental data. From this we can conclude that the description under the proposed model (3) - (5) fits better with the physics of the process.

## 5. ACKNOWLEDGMENTS

This paper is supported in part by the Presidium of RAS (projects #1.8 and 2.1).

**References**
[1] Etchegoin P G, Le Ru E C, and Meyer M. Erratum: An analytic model for the optical properties of gold. J. Chem. Phys 125 164705, 2006.
[2] Vial A., and Laroche T. Comparison of gold and silver dispersion laws suitable for FDTD simulations. Applied Physics B 93 139-143, 2008.
[3] Vial A. and Laroche T. Description of dispersion properties of metals by means of the critical points method and application to the study of resonant structures using the FDTD method. J. Phys. D: Appl. Phys. 40 7152-8, 2007.
[4] Ford G W, and Weber W H. Electromagnetic interactions of molecules with metal surfaces. Phys. Rep. 113 195-287, 1984.
[5] Rojas R, and Claro F. Theory of surface enhanced Raman scattering in colloids. J.Chem. Phys. 98 (2) 998-1006, 1993.
[6] Rakic A D, Djurisic, A B et al. Optical properties of metal films for vertical-cavity optoelectronic devices. Applied Optics 37 5271-5283, 1998.
[7] Jian-Hong Qiu, Peng Zhou, Xiao-Yong Gao et al. Ellipsometric Study of the Optical Properties of Silver Oxide Prepared by Reactive Magnetron Sputtering. J. Korean Phys. Soc., Vol. 46, pp. S269 S275, 2005.
[8] Sancho-Parramon J., Modreanu M., Bosch S., Stchakovsky M. Optical Characterization of HfO2 by

spectroscopic ellipsometry: Dispersion models and direct data inversion. Thin Solid Films, v. 516, pp.7990-7995, 2008.
[9]  Johnson P B, Christy R W. Optical constants of the noble metals. Phys. Rev. B 6, pp.4370–9, 1972.
[10] Palik E D (ed) Handbook of Optical Constants of Solids, (New York: Academic). 1985.
[11] http://www.sspectra.com/files/misc/win/SOPRA.EXE
[12] Ashcroft N. W., and Mermin N. D. Solid State Physics (Saunders College Publishing) p.10, 1976.
[13] Ehrenreich H, and Philipp H R  Optical properties of Ag and Cu. Phys. Rev. 128  1622-9 , 1962.
[14] Palagushkin A N, Prokopenko et al.  Measurement of Metal nanolayers Optical Parameters Using Surface Plasmon Resonance Method. Optical Memory and Neural Networks (Information Optics) 16  288-294, 2007.